\documentclass[9pt,twocolumn]{article}
\usepackage{arxiv}
\usepackage[utf8]{inputenc} 
\usepackage[T1]{fontenc}    
\usepackage{hyperref}       
\usepackage{microtype}      

\usepackage{amsmath}
\usepackage{amssymb}
\usepackage{rotating}
\usepackage{cuted}

\usepackage{flushend}
 \usepackage{multirow}

\title{Simultaneous measurement of phase transmission and linear or circular dichroism of an object under test}
\author{
Sergej Rothau, Norbert Lindlein  \\
Institute of Optics, Information, and Photonics\\
Friedrich-Alexander-University Erlangen-N\"urnberg (FAU)\\
Staudtstr. 7/B2, 91058 Erlangen\\
  \texttt{sergej.rothau@fau.de} \\
   \And
 Xiao Rao\\
  Master programme in advanced optical technologies (MAOT), \\
	Paul-Gordan-Str. 6, 91052 Erlangen\\
}
\begin{document} 
\maketitle
\vspace{10cm}
\begin{strip}
	\begin{abstract}	
This publication presents a novel interferometric method for the simultaneous spatially resolved analysis of an object under test regarding the phase transmission function and the magnitude and orientation of dichroism. Analogous to the classical phase shifting interferometry the measurement strategy is based on the variation of phase and polarization in an interferometer. This procedure allows to analyse simultaneously and spatially resolved the dichroic properties of the object and its impact on the phase of the incoming light in one measurement cycle. The theoretical description of the investigated methods and their experimental implementation are presented.
	\end{abstract}
\end{strip}


\section{Introduction}
The polarization-dependent absorption, or more precisely the linear and the circular dichroism is a important property of many special materials and optical elements \cite{Lin298,Krause:08}. The standart analysis of magnitude and orientation of dichroism is done by polarization characterization in transmission \cite{jones1995circular,Leidinger:15}. Especially elements with nano-structured metal layers can show some absorbing behaviour, that should be analysed parallel to its phase transmission. The classical interferometric phase analysis \cite{Malacara2007}, that can be used for that reason, can fail by inconvenient orientation of the illuminating polarization and the dichroic object, caused by low or even vanishing fringe contrast \cite{Rothau:17}.

The polarization and phase shifting interferometry method (PPSI) has been already presented in several works. It provides a sophisticated possibility for the simultaneous measurement of the phase and the polarization state of the given object wave \cite{Rothau:17} or the direct characterization of the birefringence and the phase transmission of the object under test \cite{Rothau:18}. This measurement is full-field and takes place in one measuring cycle in transmission. For the better distinction the simultaneous analysis of light waves concerning their arbitrary elliptical polarization and concerning their phase is named e-PPSI, and the analysis of the birefringence and phase transmission of the specimen is called r-PPSI.

\begin{figure}[ht]	\centering
		\includegraphics[width=0.75\columnwidth]{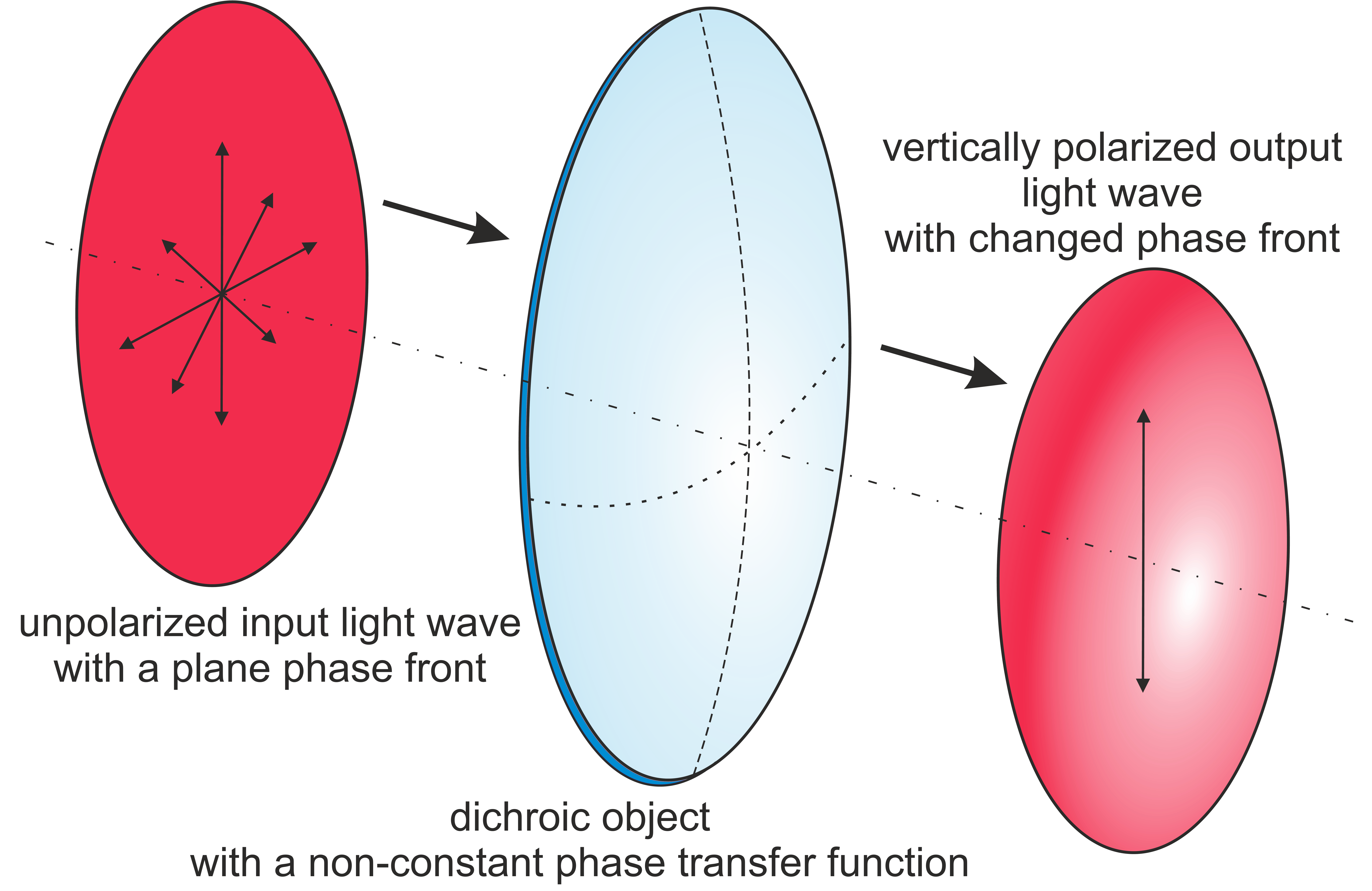}
	\caption{Scheme of the functionality of a dichroic optical element with non-constant phase transmission.}		\label{fig:intro1}
\end{figure}

The theoretical consideration of these measurement methods is based on the Jones formalism for the description of the measurement situation. Thus, it is obvious to extend this well-understood measurement procedure with the help of further Jones matrices to the measurement of other object properties, namely on the polarization-dependent absorption or more precisely on the linear and the circular dichroism (see figure \ref{fig:intro1}).

Thus, this publication describes the adaptation of the PPSI measurement strategy for the simultaneous measurement of the phase transmission ($\Phi $) and the polarization-dependent absorption or dichroism of the object under test, to the so called a-PPSI method. It is distinguished between two types of dichroism, the linear dichroism(LD) and the circular dichroism (CD). In the first case the magnitude $P_\parallel/P_\perp$ of the transmittance along two fixed principal axes and the orientation $\theta$ of these axes are measured and in the second case just the magnitudes of the transmittance $P_r/P_l$ for right- and left-handed circular polarization.

\section{Theory} \label{ref:theo}
The measurement procedure presented here is based on two-beam interferometry, so an object beam is superimposed with a reference beam with known and adjustable parameters. The object under test with the sought polarization dependent absorption and phase transmission, which is illuminated by a known incoming wave, generates the object beam. 

The observed intensity $I$, which is proportional to the square of the modulus of the total electric field $E_{OUT}$, at the exit of a two-beam interferometer with coherent and monochromatic illumination can generally be described 
by the complex-valued electric fields $\vec{E}_{O}$ (object beam) and $\vec{E}_{R}$ (reference beam) of both beams \cite{Malacara2007}: 
\begin{equation}
	I(\vec{r})\propto|\vec{E}_{OUT}(\vec{r})|^2=|\vec{E}_{O}(\vec{r})+\vec{E}_{R}(\vec{r})|^2	\label{eq:INT1}
\end{equation}

\begin{figure}[ht]	\centering
		\includegraphics[width=0.99\columnwidth]{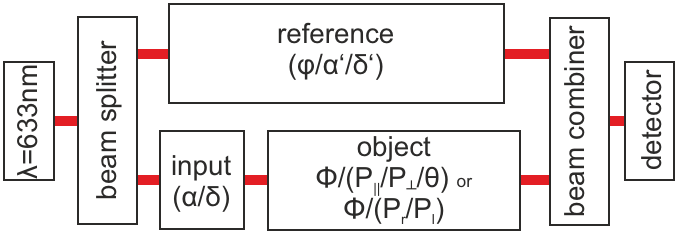}
	\caption{Scheme of the measurement setup with relevant elements and used mathematical values.}		\label{fig:setup2}
\end{figure}

In order to better understand the following theory section, it may be useful to read at first the mathematical description of the measurement procedure and the solution processes of e/r-PPSI in \cite{Rothau:17, Rothau:18}. Figure \ref{fig:setup2} shows a sketch of the measurement configuration for a-PPSI with corresponding values introduced below.

The derivation of the presented method of polarization and phase shifting interferometry is very similar to the classical PSI, but with the additional inclusion of the polarization and absorption informations. For the description of the known reference beam $\vec{E}_{R}$ with the phase-term $e^{i\varphi}$, the Jones vector with the angles $\alpha'/\delta'$, representing the polarization state, was chosen:
\begin{equation}
	\vec{E}_{R}=E_{R}\left(\begin{array}{c}\cos(\alpha')\\\sin(\alpha')e^{i\delta'}\end{array}\right)e^{i\varphi}.
	\label{eq:PPSI1}
\end{equation}

Thereby, the angle $\alpha$ gives the ratio and the angle $\delta$ the retardance between both orthogonal polarization components of the Jones vector. The same formalism with angles $\alpha/\delta$ can be used for the equation of the known incoming plane wave $\vec{E}_{IN}$ (with negligible phase term) in front of the object under test: 
\begin{equation}
\vec{E}_{IN}=E_{IN}\left(\begin{array}{c}\cos(\alpha)\\\sin(\alpha)e^{i\delta}\end{array}\right)
	\label{eq:PPSI2}
\end{equation}

To simplify the following calculations for circular and linear dichroism and without loss of generality it can be assumed that the reference and incoming waves are plane waves (so $\phi$ is just a constant phase shift between both plane waves) and have well-defined simple polarization states. In both cases the reference wave is assumed to be linearly polarized ($\delta'=0$) with adjustable orientations, given by angle $\alpha'$. The limitation of the polarization states of the incoming wave is dependent on the sort of dichroism to be measured: right-handed or left-handed circular polarization ($\alpha=\frac{\pi}{4}, \delta=\frac{\pi}{2}/\frac{3\pi}{2}$ (RCP/LCP) ) for the circular dichroism and linearly polarized light ($\delta=0$ with $\alpha$ as orientation) for the linear dichroism.

\begin{figure}[ht]	\centering
		\includegraphics[width=0.8\columnwidth]{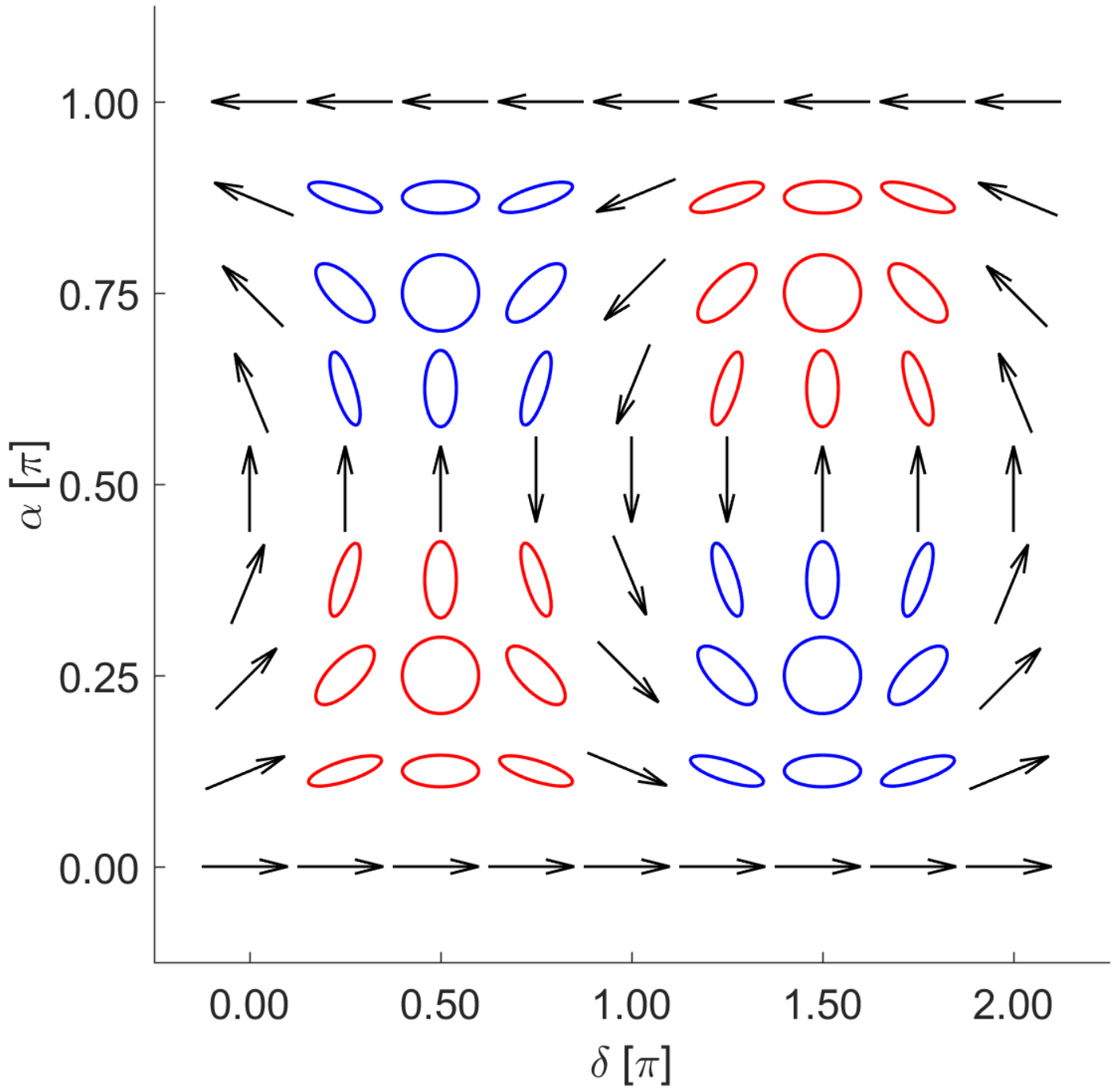}
	\caption{Polarization states as a function of $\alpha/\delta$. Red and blue ellipses stands for states
with right- and left-handed polarization, respectively.}		\label{fig:allstates}
\end{figure}

For more comprehension the space of polarization states as function of both angles alpha and delta are shown in figure \ref{fig:allstates}.

The Jones matrix of the general absorber \cite{goldstein2011polarized} for the circular (\ref{eq:PPSI3C}) and linear (\ref{eq:PPSI3L}) polarizations are used for the object description.
\begin{equation}
\hat{J}_{CD}(P_r,P_l)=\frac{1}{2}\begin{pmatrix}	P_r+P_l	&	-i (P_r - P_l) \\	i(P_r - P_l) &	P_r + P_l \end{pmatrix} 
\label{eq:PPSI3C}
\end{equation}
\begin{align}
\begin{split}
\hat{J}&_{LD}(P_\parallel,P_\perp,\theta)=\\
=&\begin{pmatrix}	P_\parallel \cos^2\theta + P_\perp \sin^2\theta	&	(P_\parallel - P_\perp) \cos\theta \sin\theta \\
													(P_\parallel - P_\perp) \cos\theta \sin\theta 			&	P_\parallel \sin^2\theta + P_\perp \cos^2\theta	\end{pmatrix}
\label{eq:PPSI3L}
\end{split}
\end{align}

The absorption of the linear polarization, therefore the linear dichroism (LD), is specified by its magnitude ($P_{\parallel/\perp}\in[0,1]$) and its orientation ($\theta \in[-\frac{\pi}{2},\frac{\pi}{2}]$). For the case of the circular dichroism (CD), so the absorption of circular polarization states, the description with only the magnitude ($P_{r/l}\in[0,1]$) is sufficient. The value 1 indicates the complete transmission and the value 0 the complete absorption of the corresponding polarization state.

The resulting light distribution $\vec{E}_{O}$ behind the object with the circular and linear dichroism can be calculated now with Eq. (\ref{eq:PPSI2}) and (\ref{eq:PPSI3C}) or (\ref{eq:PPSI3L}), respectively, to obtain the following final expression:
\begin{equation}
	\vec{E}_{O}=\hat{J}_{CD/LD}\vec{E}_{IN}(\alpha,\delta)e^{i\Phi(x,y)}.
	\label{eq:PPSI4}
\end{equation}

\noindent $\Phi(x,y)$ is its phase transmission function of the specimen. Finally both expressions of the object and reference Eq. (\ref{eq:PPSI1}) beams can be inserted into the intensity formula (\ref{eq:INT1}). The resulting formula of the observed intensity for circular and linear dichroic element are given in Eq. (\ref{eq:App2C}) and Eq. (\ref{eq:App2L}), respectively, and are functions of several parameters of the system:
\begin{align*}
\text{CD:} \quad		I(\varphi,\alpha',\delta,P_r,P_l,\Phi) \quad \text{and} \\
\quad \text{LD:} \quad	I(\varphi,\alpha',\alpha,P_\parallel,P_\perp,\theta,\Phi). 
\end{align*}

\subsection{a-PPSI: evaluation procedure for circular dichroism} 

The measurement strategy is similar to the classical PSI: by variation of the adjustable parameters different intensity distributions $I_i(\varphi_i,\delta_i,\alpha'_i)$ can be detected and subsequently used for the calculation of the unknown object values. In this case besides the reference phase angle $\varphi_i$, the orientation $\alpha'_i$ of the linear polarization of the reference beam and the angle $\delta_i$ of the incoming beam in front of the object should be varied, giving the method its name.

In order to derive the evaluation procedure, equation (\ref{eq:App2C}) can be rewritten, using trigonometric identities, as a scalar product of two vectors $\vec{Y}$ and $\vec{X_i}$:
\begin{equation}	\label{eq:VEC1C}
	I_i(\varphi_i,\delta_i,\alpha'_i)=\vec{X}_i^T(\varphi_i,\delta_i,\alpha'_i)\cdot\vec{Y}(I_0,E_r,E_l,\Phi).
\end{equation}

The six-line vector $\vec{X_i}$ has solely all known and adjustable values from equation (\ref{eq:App2C}) as entries:
\begin{equation}
	\vec{X_i} = \begin{pmatrix}
		1 \\
		\sin \delta_i \\
		\cos(\varphi_i-\alpha'_i)-\sin(\varphi_i-\alpha'_i-\delta_i)\\
		\cos(\varphi_i+\alpha'_i)+\sin (\varphi_i+\alpha'_i-\delta_i)\\
		\sin(\varphi_i-\alpha'_i)+\cos(\varphi_i-\alpha'_i-\delta_i)\\
		\sin(\varphi_i+\alpha'_i)-\cos(\varphi_i+\alpha'_i-\delta_i)
	\end{pmatrix}.
\label{eq:XCD}
\end{equation}

Analogous to the e-/r-PPSI procedures the vector $\vec{Y}$ has all unknown variables:
\begin{equation}
\vec{Y} = 
	\frac{1}{\sqrt 2} \begin{pmatrix}
		\sqrt{2} I_0 \\ \frac{1}{\sqrt{2}} (E_r^2 - E_l^2)\\
		E_R E_r \cos \Phi \\		E_R E_l \cos \Phi \\
		E_R E_r \sin \Phi \\		E_R E_l \sin \Phi
	\end{pmatrix}.
\label{eq:VecY}
\end{equation}
	
\noindent Thereby some additional auxiliary values were defined: 
\begin{itemize}
	\item products of the incoming electric field and the polarization dependent transmission coefficients:
	\begin{equation}
	E_r = E_{IN}\cdot P_r; \quad E_l=E_{IN}\cdot P_l
	\label{eq:Def1}
	\end{equation} 
	\item the definition of the mean intensity is now slightly different from the classical: 
	\begin{equation}
	I_0 = E_R^2 + \frac{1}{2}(E_r^2 + E_l^2);
	\label{eq:Def2}
	\end{equation}
	\item As an extension we define in analogy to the classical contrast formula the auxiliary visibility values separated by polarization states. These intermediate results are used later for the $P_{r/l}$ calculation in Eq. \ref{eq:Def3}:
	\begin{align} \begin{split}
		V_r = \frac{2E_R E_r}{E_R^2 + \frac{1}{2}(E_r^2 + E_l^2)}; \\
		V_l  = \frac{2E_R E_l}{E_R^2 + \frac{1}{2}(E_r^2 + E_l^2)};
	\label{eq:Def3} \end{split}
	\end{align}
\end{itemize}

For the extraction of the transmission coefficients $P_{r/l}$ from the products $E_{r/l}$ (see Eq. (\ref{eq:Def1})) a setup calibration without the object is necessary. This so called empty measurement is additionally done in a classical PSI, to get the setup phase $\Phi^e$. The subtraction of the setup phase from the measured phase eliminates the setup aberrations and delivers the pure object phase: $\Phi^{obj}=\Phi-\Phi^e$. The same procedure can be applied for PPSI measurements, too.

Beside the setup phase, the measured mean intensity $I^e_0=E_{IN}^2 + E_R^2$ and the visibility $V^e=(2 E_R E_{IN})/I_0^e$ of the empty setup with classical definitions \cite{Malacara2007} can be used for the $P_{r/l}$ calculation.
\begin{align} \begin{split}
E_R E_{IN} = \frac{V^e I_0^e}{2} = \frac{V_r I_0}{2P_r} = \frac{V_l I_0}{2P_l}; \\
 \mapsto P_r = \frac{V_r I_0}{V^e I_0^e}; \quad P_l = \frac{V_l I_0}{V^e I_0^e};
\label{eq:Def4} \end{split}
\end{align}

The following solution process is identical to the solution process described in \cite{Rothau:17}. With the vector $\vec{X_i}$ from Eq. (\ref{eq:XCD}), an auxiliary matrix $\hat{A}$ can be defined:
\begin{equation}	\label{eq:VEC4}
	 \hat{A}=\sum^N_{i=1}\hat{A_i} \qquad \text{with } \hat{A_i}=\vec{X_i}\vec{X_i}^T.
\end{equation}

That allows to rewrite formula (\ref{eq:VEC1C}) in its final form: 
\begin{equation}	\label{eq:VEC5}
	\sum^N_{i=1}\vec{X_i}I_i=\sum^N_{i=1}\vec{X_i}\vec{X_i}^T\vec{Y}=\hat{A}\vec{Y}.
\end{equation}
\\
\noindent 
The last step consists of finding combinations of $(\varphi_i/\delta_i/\alpha'_i)$ to make the matrix $\hat{A}$ invertible, delivering the final expression:
\begin{equation}	\label{eq:VEC6}
	\vec{Y}=\hat{A}^{-1}\sum^N_{i=1}\vec{X}_iI_i.
\end{equation}

If a suitable combination of the steps is used, the matrix $\hat{A}^{-1}$ and vectors $X_i$ are known and the corresponding intensities $I_i$ can be detected. This implies, that the unknown vector $\vec{Y}$ can be calculated with the equation (\ref{eq:VEC6}). Therefore all unknown values can be obtained from the calculated vector $\vec{Y}$ with equations (\ref{eq:calc1})-(\ref{eq:calc4}).
\begin{equation}
I_0 = Y_1; 
\label{eq:calc1}
\end{equation}

\begin{equation}
\Phi_1 =\arctan \left(\frac{Y_5}{Y_3}\right);  \quad \Phi_2=\arctan \left(\frac{Y_6}{Y_4}\right);
\label{eq:calc2}
\end{equation}

\begin{equation}
V_r = \frac{2\sqrt{2} \cdot \sqrt{Y_3^2 + Y_5^2}}{Y_1}; \quad
			V_l = \frac{2\sqrt{2} \cdot \sqrt{Y_4^2 + Y_6^2}}{Y_1}; 
			\label{eq:calc3}
\end{equation}

\begin{equation}
P_r = \frac{2\sqrt{2} \sqrt{Y_3^2 + Y_5^2}}{V^e I_0^e};				
\quad P_l = \frac{2\sqrt{2} \sqrt{Y_4^2 + Y_6^2}}{V^e I_0^e}; 
\label{eq:calc4}
\end{equation}

The phase distribution $\Phi$ can be evaluated on two ways ($\Phi_{1/2}$), depending on the dichroic property of the object: $P_r\approx 0 \mapsto \Phi_2$ and $P_l\approx 0 \mapsto \Phi_1$. For the calculation of pure transmission coefficients $P_{r/l}$ the results $V^e$ and $I_0^e$ from the empty measurement should be taken into account, as described before.

\subsection{a-PPSI: evaluation procedure for linear dichroism} 
The evaluation procedure for the linear dichroism is similar to the previous one. The intensity distributions $I_i(\varphi_i,\alpha_i,\alpha'_i)$ from Eq. (\ref{eq:App2L}) is rewritten in a scalar product of two vectors $\vec{Y}$ and $\vec{X_i}$:
\begin{equation}	\label{eq:VEC1L}
	I_i(\varphi_i,\alpha_i,\alpha'_i)=\vec{X}_i^T(\varphi_i,\alpha_i,\alpha'_i)\cdot\vec{Y}(I_0,E_\parallel,E_\perp,\theta,\Phi).
\end{equation}

The separation was again for known and unknown values, so the nine-lines vector $\vec{X}_i$ includes all known and adjustable quantities:
\begin{equation} \label{eq:XLD}
	\vec{X_i}  = \begin{pmatrix}			1 \\  			
	\cos (2\alpha_i) \\	\sin (2\alpha_i) \\
	\cos(\alpha_i - \alpha'_i) \cos\varphi_i \\	\cos(\alpha_i - \alpha'_i) \sin\varphi_i \\
	\cos(\alpha_i + \alpha'_i) \cos\varphi_i \\	\cos(\alpha_i + \alpha'_i) \sin\varphi_i \\
	\sin(\alpha_i + \alpha'_i) \cos\varphi_i \\	\sin(\alpha_i + \alpha'_i) \sin\varphi_i	\end{pmatrix}.
\end{equation}

On the other hand the nine-line vector $\vec{Y}$ has all unknown values from equation (\ref{eq:App2L}) as entries:
\begin{equation}
	\vec{Y}= \begin{pmatrix}	
	I_0\\ \frac{1}{2}(E_\parallel^2 - E_\perp^2) \cos (2\theta) \\		\frac{1}{2}(E_\parallel^2 - E_\perp^2)\sin (2\theta) \\
																									E_R \cos\Phi (E_\parallel + E_\perp) \\	E_R \sin\Phi (E_\parallel + E_\perp) \\
																									E_R \cos\Phi \cos2\theta (E_\parallel - E_\perp) \\	E_R\sin\Phi \cos2\theta (E_\parallel - E_\perp) \\
																									E_R \cos\Phi \sin2\theta (E_\parallel - E_\perp) \\	E_R \sin\Phi \sin2\theta (E_\parallel - E_\perp)			\end{pmatrix}.
\end{equation}

Also the auxiliary values similar to the previous section (see Eq. (\ref{eq:Def1})-(\ref{eq:Def3})) are defined:
\begin{equation}
	E_\parallel = E_{IN} \cdot P_\parallel; \quad		E_\perp = E_{IN} \cdot P_\perp;	
\end{equation}

\begin{equation}
	I_0 = E_R^2 + \frac{1}{2}(E_\perp^2 + E_\parallel^2);	
\end{equation}

\begin{equation}
	V_\parallel = \frac{2E_R E_\parallel}{E_R^2 + \frac{1}{2}(E_\parallel^2 + E_\perp^2)}; \quad	V_\perp = \frac{2E_R E_\perp}{E_R^2 + \frac{1}{2}(E_\parallel^2 + E_\perp^2)};
\end{equation}

The mean intensity $I_0^e$ and the visibility $V^e$ from the additional empty measurement allows the calculation of the transmission coefficients $P_{\parallel/\perp}$ analogue to Eq. (\ref{eq:Def4}):
\begin{align} \begin{split}
E_R E_{IN} = \frac{V^e I_0^e}{2} = \frac{V_\parallel I_0}{2P_\parallel} = \frac{V_\perp I_0}{2P_\perp}; \\ 
\mapsto P_\parallel = \frac{V_\parallel I_0}{V^e I_0^e}; \quad P_\perp = \frac{V_\perp I_0}{V^e I_0^e}
\label{eq:Def5}  \end{split} \end{align} 

The last step of the evaluation is the calculation of the auxiliary matrix $\hat{A}$ (see Eq. (\ref{eq:VEC4})) and searching for the right combinations of $(\varphi_i,\alpha_i,\alpha'_i)$ for the inverting to $\hat{A}^{-1}$ (see Eq. (\ref{eq:VEC6})). With this inverted matrix and the additional empty measurement results all unknown values can be calculated:

\begin{equation}
I_0 = Y_1		
\end{equation}

\begin{equation}
\Phi= \arctan\left(\frac{Y_5}{Y_4}\right)
\end{equation}
	
\begin{equation}
\theta=\frac{1}{2}\arctan \left(\frac{Y_3}{Y_2}\right)
	\label{eq:thCD}
\end{equation}
			
\begin{equation}
V_{\parallel} = \frac{\sqrt{Y_4^2 +Y_5^2}+\sqrt{Y_6^2 + Y_7^2 + Y_8^2 + Y_9^2}}{Y_1} 
\end{equation}

\begin{equation}
V_{\perp} = \frac{\sqrt{Y_4^2 +Y_5^2}-\sqrt{Y_6^2 + Y_7^2 + Y_8^2 + Y_9^2}}{Y_1}
\end{equation}

\begin{equation}
P_{\parallel/\perp} = \frac{\sqrt{Y_4^2 +Y_5^2}\pm\sqrt{Y_6^2 + Y_7^2 + Y_8^2 + Y_9^2}}{V^e I_0^e}; 
\end{equation}

Equation (\ref{eq:thCD}) gives no result if the values of $E_\perp$ and $E_\parallel$ are equal. This special case means that the object under test absorbs the light without any polarization dependence like a standard density filter with $P=P_\perp=P_\parallel$ and undefined preferential direction $\theta$. 

Furthermore, the phase $\Phi$ can be calculated on alternative ways by $\arctan(Y_7/Y_6)$ or $\arctan(Y_9/Y_8)$ with the restrictions, that $E_\perp\neq E_\parallel$, $\theta\neq \pi/4$ or $\theta\neq 0$, respectively. If these conditions are fulfilled, this additional phase results can be taken for the consistence test of a measurement.
	
\section{Measurement algorithms} \label{ref:alg}
There is theoretically an infinite number of different combinations of the phase and polarization steps for both measurement situations making matrix $\hat{A}$ from Eq. (\ref{eq:VEC4}) defined as a product of vector $\vec{X}$ from Eq. (\ref{eq:XCD}) and Eq. (\ref{eq:XLD}), respectively, invertible and thus can be used for the measurement. For practical reasons, step values are limited by several assumptions, like preferably small number of equidistant steps, variation range from $0$ to $\pi$ or $2\pi$ and so on. 

In principle, in these measuring methods, the phase is shifted and additionally the polarization in both arms is also varied. As already mentioned in section \ref{ref:theo}, the allowed polarization states of the light waves in the interferometer arms are limited:
\begin{itemize}
	\item CD: $(\alpha=\pi/4,$ $\delta\in \{\pi/2,3\pi/2\})$ / $(\delta'=0,$ $\alpha' \in [0,2\pi[)$
	\item LD: $(\alpha\in [0,2\pi[,$ $\delta=0)$ / $(\delta'=0,$ $\alpha' \in [0,2\pi[)$
\end{itemize}

The possible algorithms can be summarized in different groups sorted e.g. by the variation strategy of the polarization states in both arms. In the next two subsections, several examples of different algorithms for the measurement of circular and linear dichroism are presented. For a more qualitative characterization of the algorithms, the determinant $\det(\hat{A})$ and condition number $\kappa(\hat{A})=\left\|\hat{A}\right\| \left\|\hat{A}^{-1}\right\|$ of the corresponding matrix $\hat{A}$ are given. Thereby the condition number $\kappa$ is calculated with the standard euclidean matrix norm. Large values of the determinant and small condition numbers are usually a convincing criterion for a robust algorithm.

The range of the phase shifting steps can be either not averaging or averaging, like in the well-known 4 or 5 step algorithms of PSI, respectively \cite{Malacara2007}. In the case of averaging algorithms, the first frame with the phase value $\varphi=0$ should theoretically be identical with the last frame with phase value $\varphi=2\pi$. For the classical averaging 5 step algorithm this results in more robustness against linear phase stepping errors \cite{Schwider:83}.

\subsection{Algorithms for circular dichroism}
Like in previous PPSI versions, there is a huge number of different algorithms: three different examples are presented in table \ref{tab:ALG1} and discussed in detail.  In both tables \ref{tab:ALG1}-\ref{tab:ALG2} the arrow pictograms symbolize the polarization direction given by the corresponding angle pairs: $(\alpha/\delta)$ and $(\alpha'/\delta')$.

The methods with the successive variation of the polarization in each arm are named ($N,M,L$), where $N$ designate the number of phase frames, $M$ the number of polarization states of the incoming light and $L$ the number of state of polarization used in the reference arm. The methods with simultaneous variation of polarization in both arms are named ($N/M$), where N is the number of phase steps and M the number of polarization combinations.
\begin{table}[ht]	
\caption{Examples of algorithms for the simultaneous measurement of the phase transmission and circular dichroism of the object under test; $N_{t}$ gives the total number of frames used of current method;} 
		\begin{center}       	\vspace{-0.2cm}
		\begin{tabular}{c|c|l|l|l} 
			\hline
			\rule[-1ex]{0pt}{3.5ex} & $N_{t}$ & $\varphi_i$ & $\delta_i$ & $\alpha'_i$ \\
			\hline
			\hline
			\rule[-1ex]{0pt}{3.5ex} 
			\parbox[t]{2mm}{\multirow{2}{*}{\rotatebox[origin=c]{90}{$5^*/2/1$}}}
			& 10 & $\{0, \frac{\pi}{2}, \pi, \frac{3\pi}{2}, 2\pi\} $&$\{\frac{\pi}{2},\frac{3\pi}{2}\}$&$\{0 \}$ \\
			&&& $\{\circlearrowright,\circlearrowleft\}$ & $\{\rightarrow\}$ \\
			\cline{2-5}
			& \multicolumn{4}{l}{$\det(\hat{A})=8.0e+05; \quad \kappa(\hat{A})=2.29;$} \\
			\multicolumn{5}{l}{Each 5 phase steps $\varphi_i$ for the two pol. values $\delta_i$} \\
			\hline \hline
			\rule[-1ex]{0pt}{3.5ex} 
			\parbox[t]{2mm}{\multirow{2}{*}{\rotatebox[origin=c]{90}{$3/2/1$}}}
			& 6 & $\{0, \frac{2\pi}{3}, \frac{4\pi}{3}\}$ &$\{\frac{\pi}{2},\frac{3\pi}{2}\}$&$\{0\}$\\ 
			&&& $\{\circlearrowright,\circlearrowleft\}$& $\{\rightarrow\}$ \\
			\cline{2-5}
			& \multicolumn{4}{l}{$\det(\hat{A})=4.7e+04; \quad \kappa(\hat{A})=1;$} \\
			\multicolumn{5}{l}{Each 3 phase steps $\varphi_i$ for the two pol. values $\delta_i$} \\
			\hline \hline
			\rule[-1ex]{0pt}{3.5ex} 
			\parbox[t]{2mm}{\multirow{2}{*}{\rotatebox[origin=c]{90}{$3/2A$}}}
			& 6 & $\{0, \frac{2\pi}{3}, \frac{4\pi}{3}\}$ &$\{\frac{\pi}{2},\frac{3\pi}{2}\}$&$\{0, \frac{\pi}{2}\}$\\ 
			&&&$ \{\circlearrowright,\circlearrowleft\}$& $\{\rightarrow,\uparrow\}$ \\
			\cline{2-5}
			& \multicolumn{4}{l}{$\det(\hat{A})=2.4e+06; \quad \kappa(\hat{A})=1;$} \\
			\multicolumn{5}{l}{pol. variation in both arms simult.: ($\delta_1/\alpha'_1$), ($\delta_2/\alpha'_2$)} \\
			\hline \hline
			\end{tabular} \label{tab:ALG1}
	\end{center}
\end{table}

The first two presented examples ($5^*/2/1$) and ($3/2/1$) are working with just one polarization state in the reference, namely the horizontally linearly polarized beam. The phase shifting series encompass five phase averaged and three not phase averaged frames, respectively, and are done for two polarization states of the incoming light: the right and left circular polarization. 

By the third example ($3/2A$) from the table \ref{tab:ALG1} the polarization variation is done in both arms simultaneously, that means the first phase shifting with three phase steps is done for the combination of right circular polarized incoming light and horizontally linearly polarized reference light. Before the second phase shifting is applied, the polarization states are varied to left circular and vertically linear, respectively.

\subsection{Algorithms for linear dichroism}
In this subsection some algorithms for the analysis of linear dichroism are presented in table \ref{tab:ALG2} and discussed. 

The polarization variation of the first presented algorithm ($3/3/2$) in table \ref{tab:ALG2} is done after each phase shifting series successively in each arm. So in total six different combinations of polarizations states are used for this measurement.

The next two examples ($3/3B$) and ($3/3C$) are working with nine frames, that is the minimal allowed number of frames given by the nine-lines structure of the $\vec{X}$ vector. These two algorithms are using three phase shifting series for three different polarization combinations, that are adjusted simultaneously in both arms.
\begin{table}[ht]	
\caption{Examples of algorithms for the simultaneous measurement of the phase transmission and linear dichroism of the object under test; $N_{t}$ gives the total number of frames used of current method; } 
		\begin{center}       	\vspace{-0.2cm}
		\begin{tabular}{c|c|l|l|l} 
			\hline
			\rule[-1ex]{0pt}{3.5ex}  $N_{t}$ & $\varphi_i$ & $\alpha_i$ & $\alpha'_i$ \\
			\hline
			\hline
			\rule[-1ex]{0pt}{3.5ex} 
				\parbox[t]{2mm}{\multirow{2}{*}{\rotatebox[origin=c]{90}{$3/3/2A$}}}
			& 18  & $\{0, \frac{2\pi}{3}, \frac{4\pi}{3}\} $&$\{0, \frac{\pi}{3}, \frac{2\pi}{3} \}$&$\{0, \frac{\pi}{2}\}$ \\
			&&&  \{$\rightarrow$,\rotatebox[origin=c]{60}{$\rightarrow$},\rotatebox[origin=c]{120}{$\rightarrow$} \} & $\{\rightarrow,\uparrow\}$ \\
			\cline{2-5}
			&  \multicolumn{4}{l}{$\det(\hat{A})=1.2e+07; \quad \kappa(\hat{A})=4;$} \\
			\multicolumn{5}{l}{polarization variation in each arm successively} \\
			\hline \hline
			\rule[-1ex]{0pt}{3.5ex} 
			\parbox[t]{2mm}{\multirow{2}{*}{\rotatebox[origin=c]{90}{$3/2B$}}}
			& 9 & $\{0, \frac{2\pi}{3}, \frac{4\pi}{3}\} $ &$\{0,\frac{\pi}{3},\frac{2\pi}{3}\}$& $\alpha=\alpha'$\\ 
			&&& \{$\rightarrow$,\rotatebox[origin=c]{60}{$\rightarrow$},\rotatebox[origin=c]{120}{$\rightarrow$} \}& \{$\rightarrow$,\rotatebox[origin=c]{60}{$\rightarrow$},\rotatebox[origin=c]{120}{$\rightarrow$} \} \\
			\cline{2-5}
			&  \multicolumn{4}{l}{$\det(\hat{A})=9.5e+04; \quad \kappa(\hat{A})=4;$} \\
			\multicolumn{5}{l}{pol. var. in both arms simult.:  ($\alpha_1/\alpha'_1$), ...,  ($\alpha_3/\alpha'_3$)}\\
			\hline \hline
			\rule[-1ex]{0pt}{3.5ex} 
			\parbox[t]{2mm}{\multirow{2}{*}{\rotatebox[origin=c]{90}{$3/2C$}}}
			& 9 & $\{0, \frac{2\pi}{3}, \frac{4\pi}{3}\} $ &$\{0,\frac{\pi}{3},\frac{2\pi}{3}\}$&$\{0,\frac{\pi}{4},\frac{\pi}{2}\}$\\ 
			&&& \{$\rightarrow$,\rotatebox[origin=c]{60}{$\rightarrow$},\rotatebox[origin=c]{120}{$\rightarrow$} \}& $\{\rightarrow,\rotatebox[origin=c]{45}{$\rightarrow$},\uparrow\}$ \\
			\cline{2-5}
			&  \multicolumn{4}{l}{$\det(\hat{A})=9.5e+04; \quad \kappa(\hat{A})=7.21;$} \\
			\multicolumn{5}{l}{pol. var. in both arms simult.:  ($\alpha_1/\alpha'_1$),...,  ($\alpha_3/\alpha'_3$)} \\
			\hline \hline
			\end{tabular} \label{tab:ALG2}
	\end{center}
\end{table}
	
\section{Experimental setup} \label{sec:STP}
\begin{figure}[ht]	\centering
		\fbox{\includegraphics[width=0.99\linewidth]{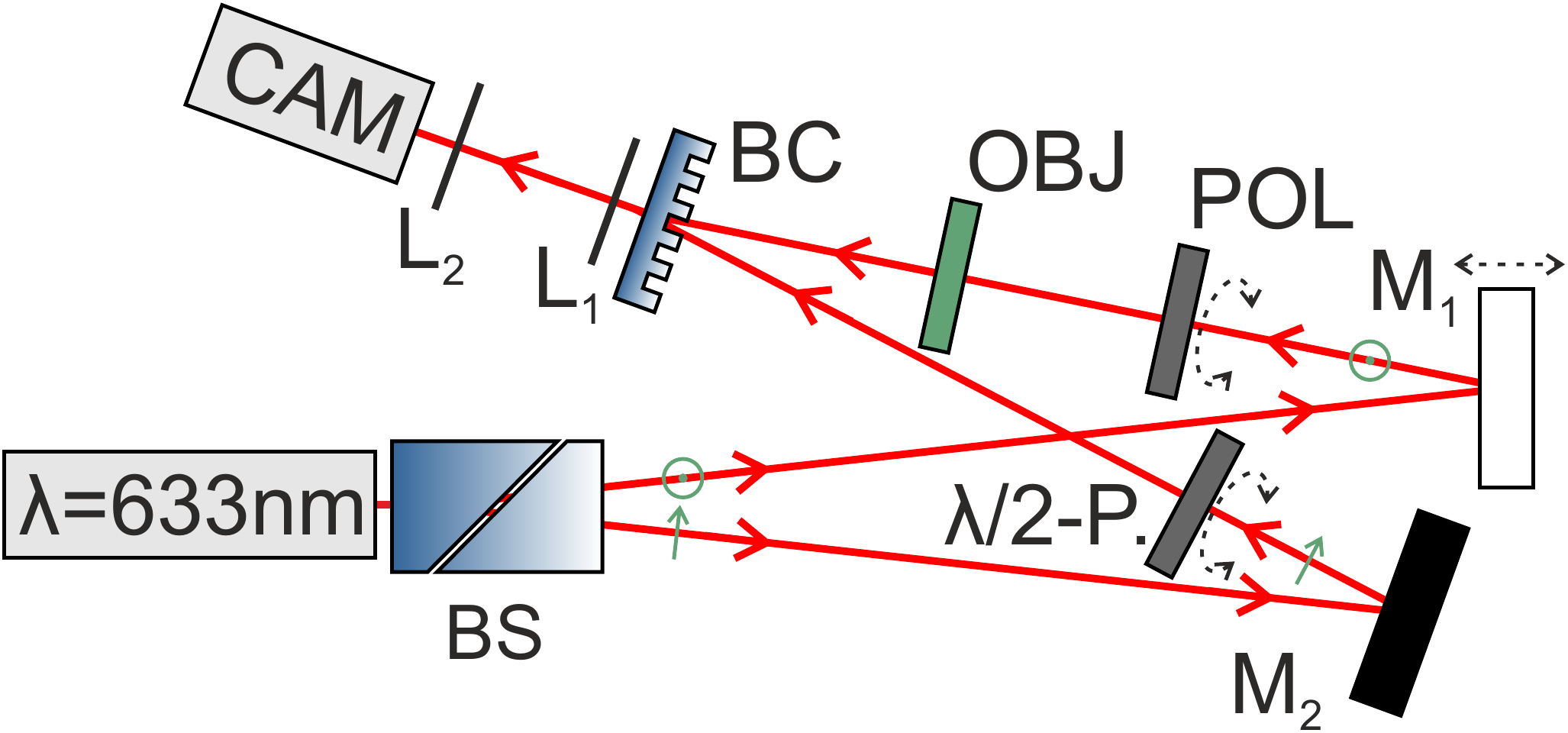}}
		\caption{Experimental PPSI-setup as a folded Mach-Zehnder-interferometer.}	\label{fig:setup1}
\end{figure}

The measurement setup is identical to the already presented setup in \cite{Rothau:17}, it is the folded Mach-Zehnder-interferometer (see Fig. \ref{fig:setup1}). 

The unusual geometry is caused by the small angles of incidence on the two mirrors, for the suppression of the influence of the reflections on the polarization states, as the Fresnel coefficients indicate. The intensity of the linearly polarized illumination with the wavelength of $\lambda=633nm$ can be adjusted by the combination of a fixed polarizer and a rotatable half-wave plate (not explicitly shown in Fig. \ref{fig:setup1}). A Wollaston prism used as beam splitter ($BS$) splits the light into its vertical and horizontal linear components. The relative phase shift between the arms is done by a movement of the mirror ($M_1$) by the piezo actuator. 

As already mentioned in section \ref{ref:theo} the reference light is limited to be linearly polarized with variable orientations given by angle $\alpha'$. This polarization rotation is done by rotatable polarizing elements: a half-wave plate ($\lambda/2-P.$). Depending on the sort of dichroism the polarizing element ($POL$) in front of the object under test ($OBJ$) is quarter- or half-wave-plate for a circular or linear dichroism, respectively. 

The object beam with the corresponding polarization transmitted through the specimen is recombined with the reference beam with a specially designed Ronchi grating ($BC$). The influence of the grating parameters such as the period, duty cycle, and etching depth on TE and TM polarization have been investigated via a rigorous diffraction method regarding diffraction efficiency and phase retardation. The grating is designed to conserve any polarization patterns present in the incident beams, thereby the angle between the incoming beams is $15^\circ$.

For the detection of the interferograms a camera ($CAM$) with an imaging telescope ($L_1$ and $L_2$) behind the beam combiner is used.

The PSI and other PPSI methods can be used in such setup by keeping the polarizations in both arms or just in the object arm constant, while phase or phase and polarization of the reference beam are shifted, respectively. That allows to compare different PPSI results with each other and with PSI results directly without moving the object and without any setup modification. 

The maximum size of the measurement fields was limited by the used optical components and was about 3cm. The metrological characteristic for the setup is presented by the following average values for repeatability and reproducibility:
\begin{itemize}
	\item repeatability: \\ \quad \quad		
	$\sigma_{\Phi}=0.01 \pi$; $\sigma_{P}=0.005$; (LD: $\sigma_{\theta}=0.01 \pi$);
	\item reproducibility: \\ \quad \quad
	$\sigma_{\Phi}=0.04 \pi$; $\sigma_{P}=0.01$; (LD: $\sigma_{\theta}=0.02 \pi$);
\end{itemize}
Thereby, repeatability is the rms value of the difference of multiple measurement results, made in succession without any changing on the setup. The reproducibility gives the rms value of the difference of multiple results, where the specimen has been removed and was inserted again, and the setup was realigned. The accuracy of these statistical characteristics of the measurement procedure increase with the increasing numbers of measurements, so at least 10 or more results should be taken into account.

\section{Experimental results}		
In the two following subsections after a short description of the function of the objects under test the distributions of the measured quantities are shown and discussed.

\subsection{CD measurement results}
For the test of the CD-analysis a simulated circular dichroic object as a combination of two quarter wave plates and a linear polarizer in between was used. The relative orientation of these three elements is crucial. For the ideal right $(P_r=1,P_l=0)$ and left $(P_l=1, P_r=0)$ dichroic absorber the first quarter wave plate is placed under angle $\alpha$, the following polarizer under angle $\alpha+45^\circ$ for right circular and $\alpha-45^\circ$ for left circular dichroism and the second quarter wave plate under $\alpha+90^\circ$. 

This can be easily proved with Jones calculus (e.g. $\alpha=0$ and $P_r=1,P_l=0$). The final Jones matrix (compare with Eq. (\ref{eq:PPSI3C})) is a product of the three elements:
\begin{align}
\begin{split}
\hat{J}_{CD}(P_r=1,P_l=0)&=\hat{J}_{\frac{\lambda}{4}}(90^\circ)\hat{J}_{linPol}(+45^\circ) \hat{J}_{\frac{\lambda}{4}}(0^\circ)\\
&=\frac{1}{2}\begin{bmatrix}1 & -i \\ i & 1\end{bmatrix}.
\end{split}
\end{align}

\begin{figure}[ht]
			\centering
				\includegraphics[width=0.49\columnwidth]{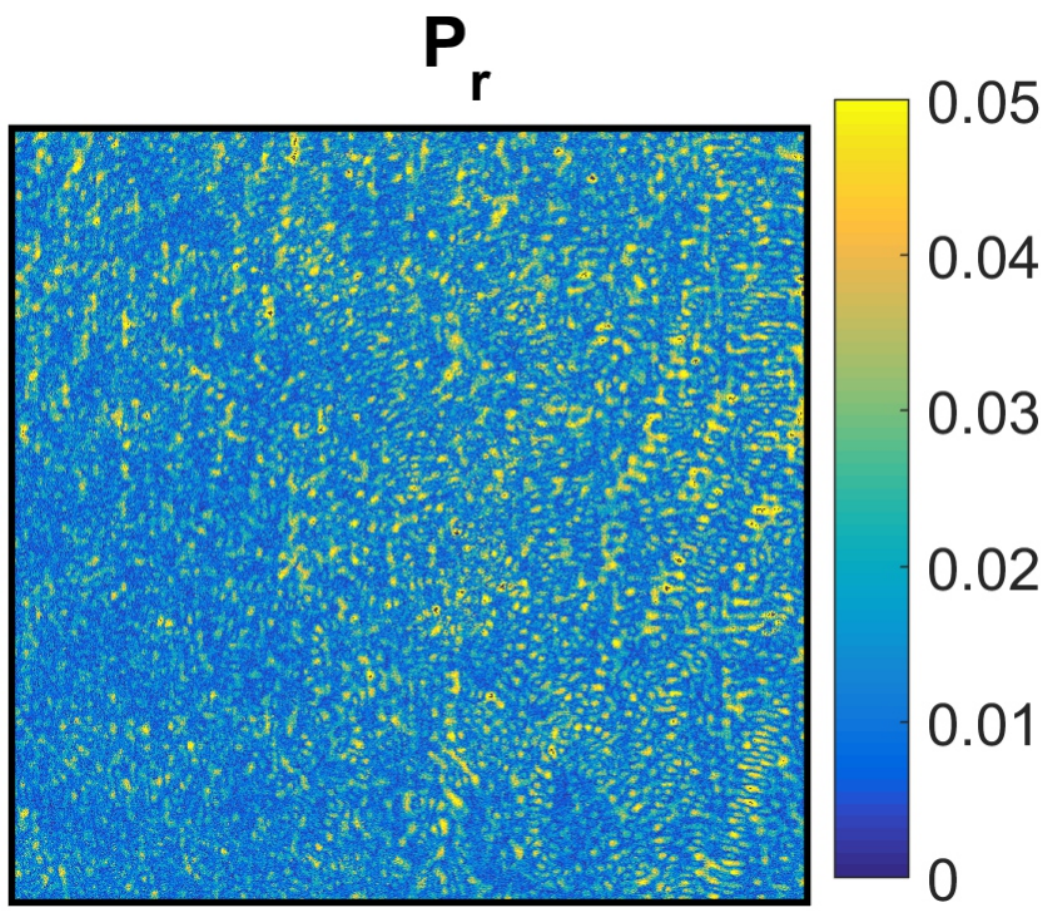} 
				\includegraphics[width=0.49\columnwidth]{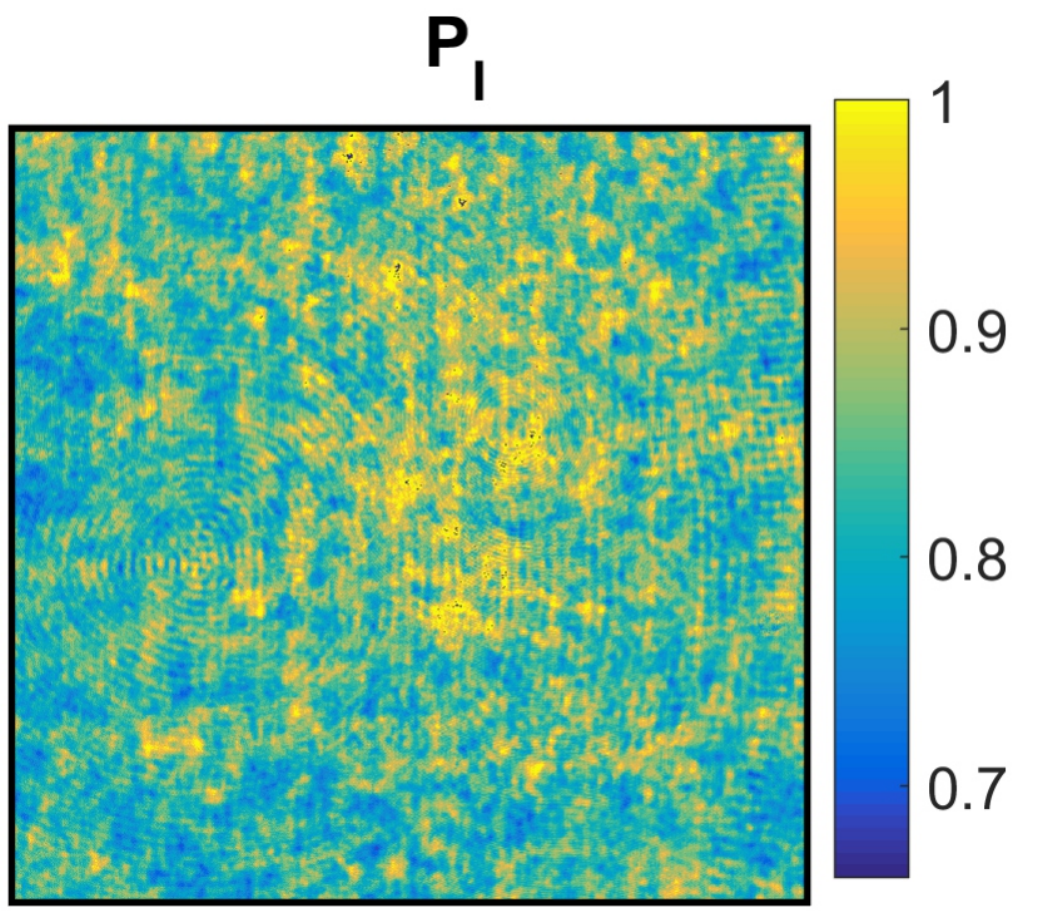} \\
				\includegraphics[width=0.49\columnwidth]{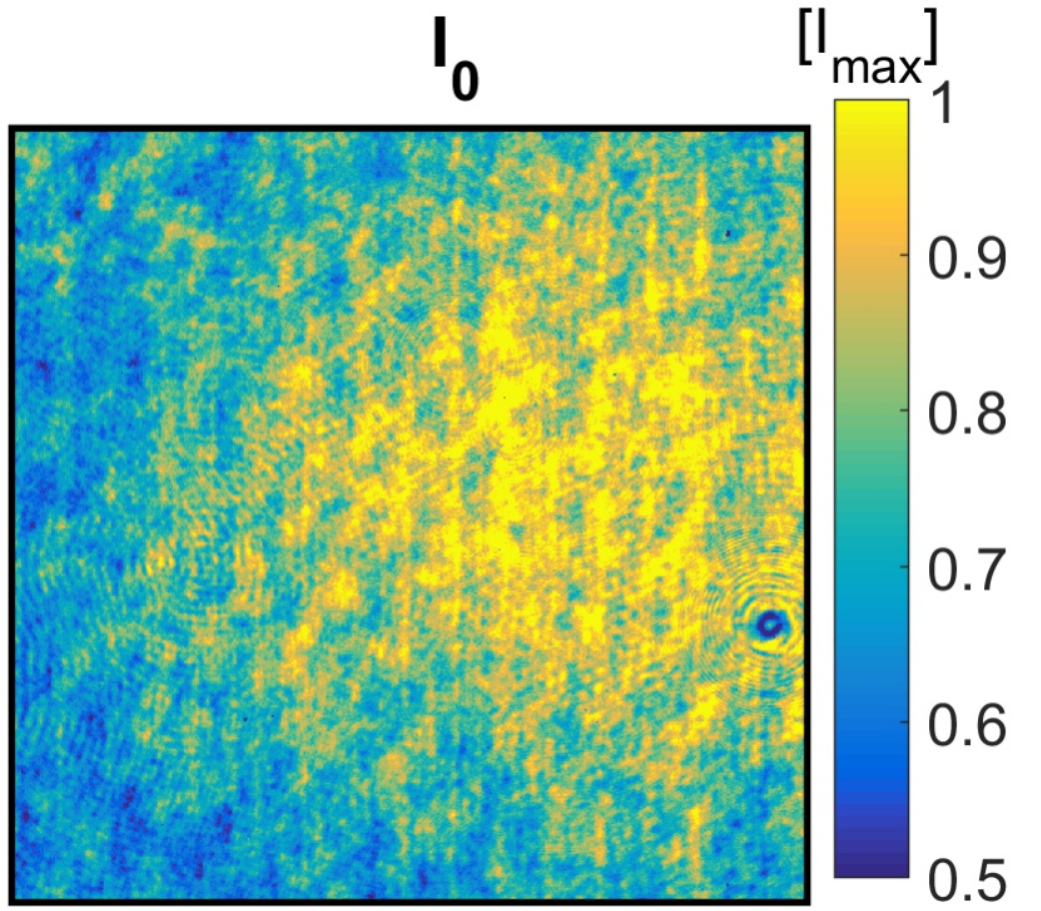}
				\includegraphics[width=0.49\columnwidth]{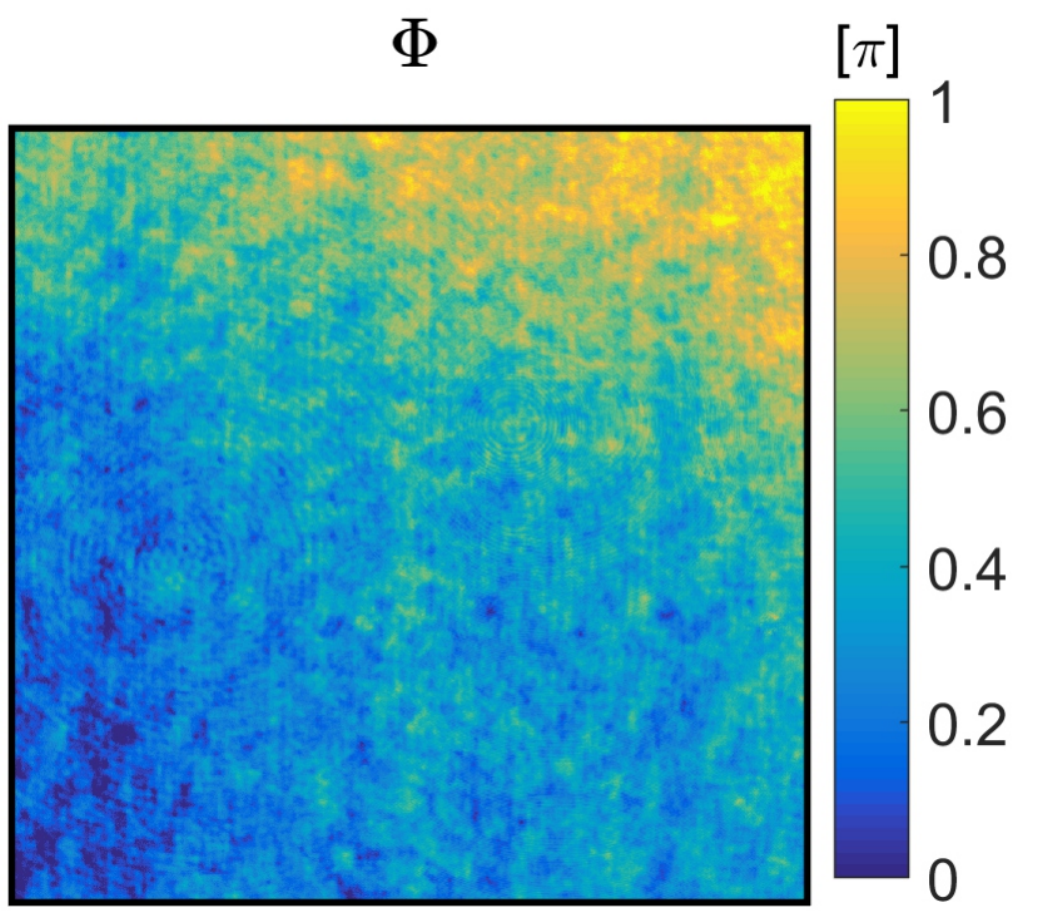}
			\caption{Measurement results of the effective left circular dichroic sample with a-PPSI $(5^*/2/1)$ algorithm: transmission coefficients $P_{r/l}$, mean intensity $I_0$ and phase front $\Phi$.}
			\label{fig:RES2a}
		\end{figure}

Figure \ref{fig:RES2a} presents the results of a a-PPSI measurement of such an effectively left circular dichroic object with a $(5^*/2/1)$ algorithm from table \ref{tab:ALG1}.
		
The measured value of $P_r$ is nearly constant zero and of $P_l$ is maximal, so the results indicate a left circular dichroism of the specimen. The phase distribution $\Phi$ has a small tilt over a half wavelength as a result of the combination of three polarising elements to the effective dichroic absorber. The mean intensity $I_0$ is very similar to the mean intensity measured in the LD-analysis (see Fig. \ref{fig:RES1a}), due to the unchanged illumination in the setup.

\subsection{LD measurement results}

\begin{figure}[ht]
			\centering
				\includegraphics[width=0.75\columnwidth]{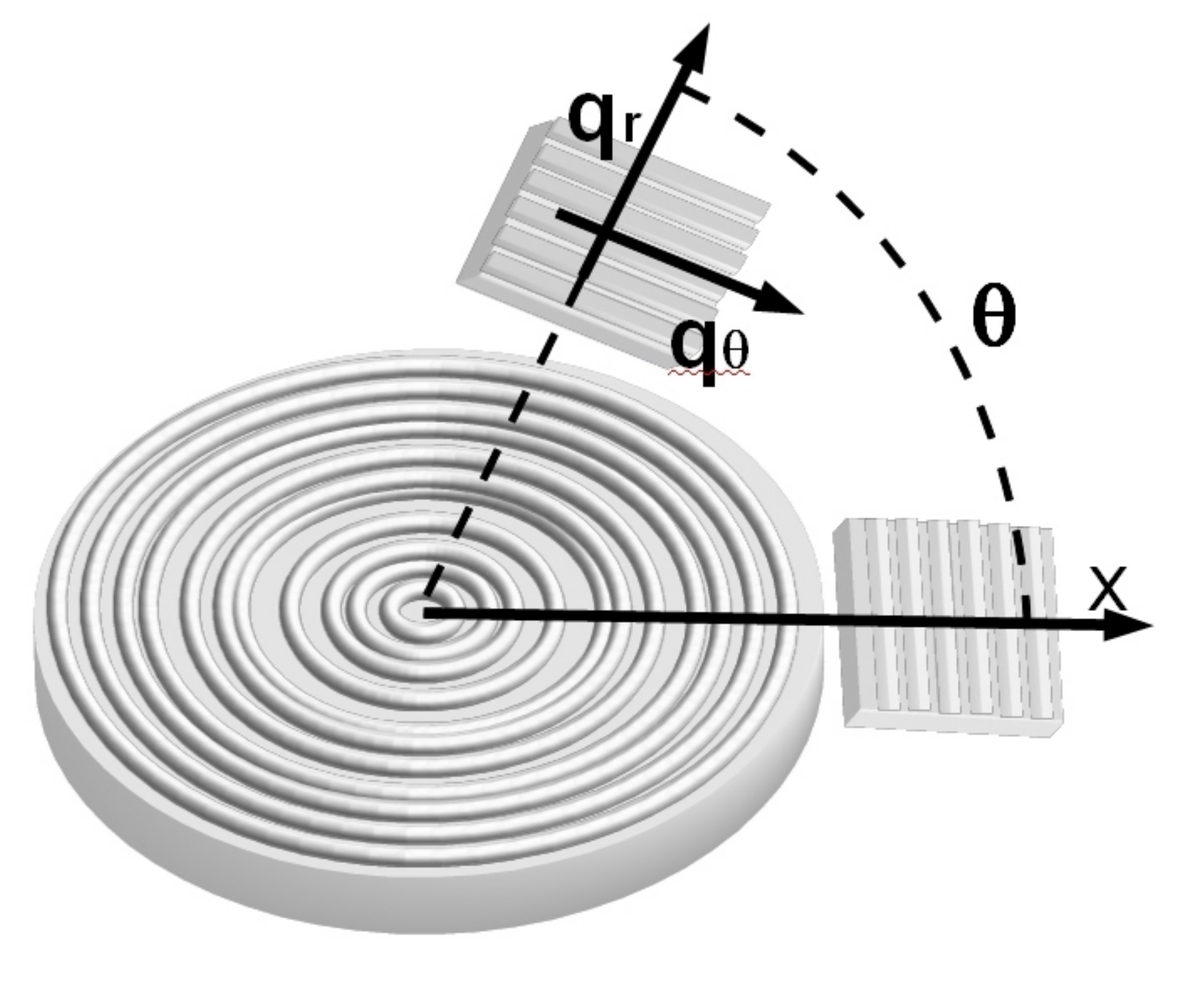}
			\caption{Geometry of concentric ring polarizer as space-variant local polarizer.}
			\label{fig:RES0a}
\end{figure}

The device under test was a radial polarizer made from a sub-wavelength aluminum circular grating on a glass substrate for the wavelength $\lambda= 633nm$ \cite{Ghadyani:11}. This polarizing element is normally used to convert the circular input polarization into so called radial polarization where the light is locally linearly polarized parallel to the local grating vector of the circular grating. A concentric ring metal grating with sub-wavelength period is principally a space-variant polarizer with its transmission axis oriented in radial direction. It can be locally described by the Jones matrix of a standard polarizer with its transmission axis rotated by azimuthal angle $\theta$ with respect to the x axis (see fig. \ref{fig:RES0a}).

\begin{figure}[ht]
			\centering
				\includegraphics[width=0.49\columnwidth]{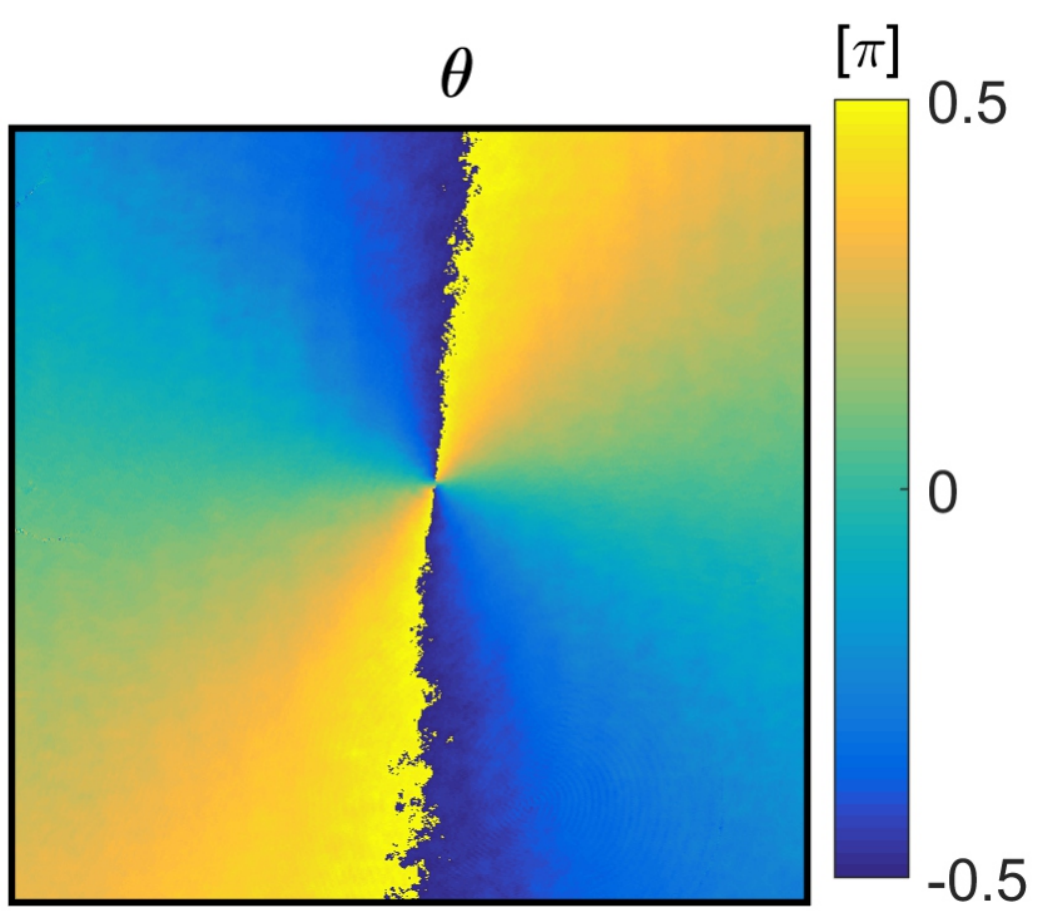} \\
				\includegraphics[width=0.49\columnwidth]{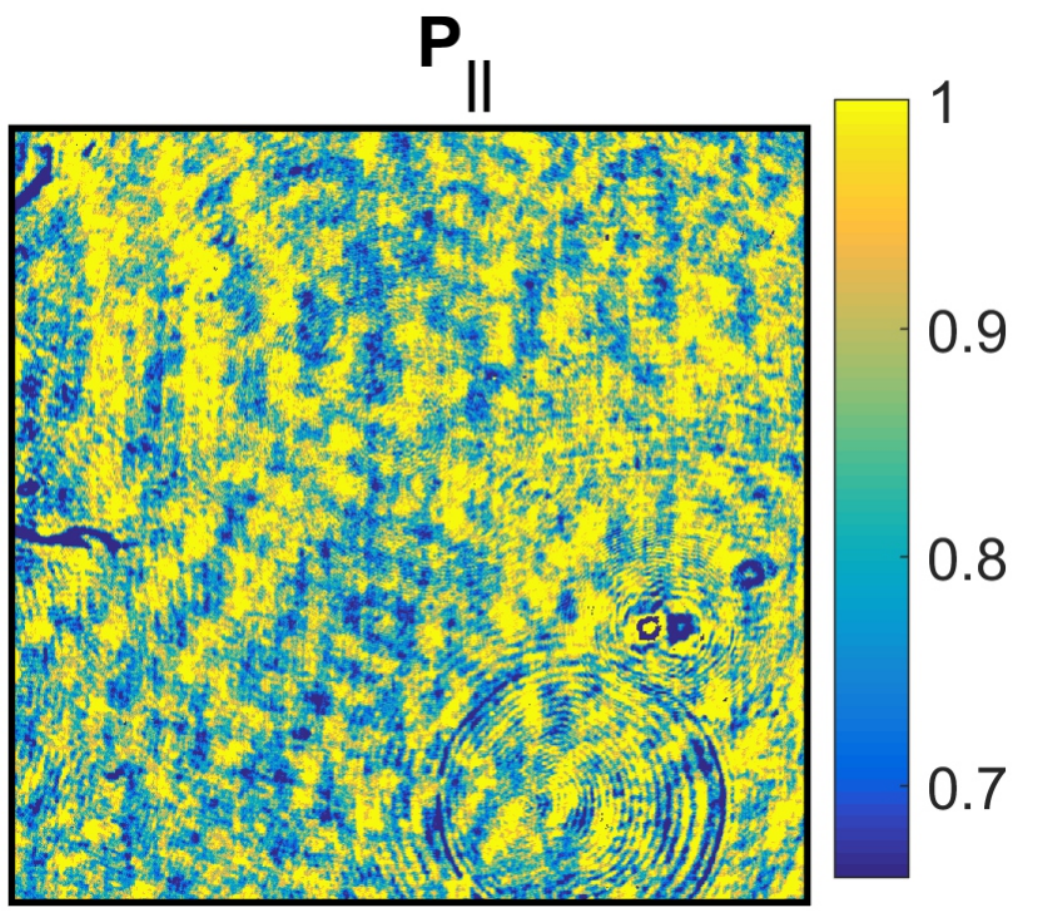} 
				\includegraphics[width=0.49\columnwidth]{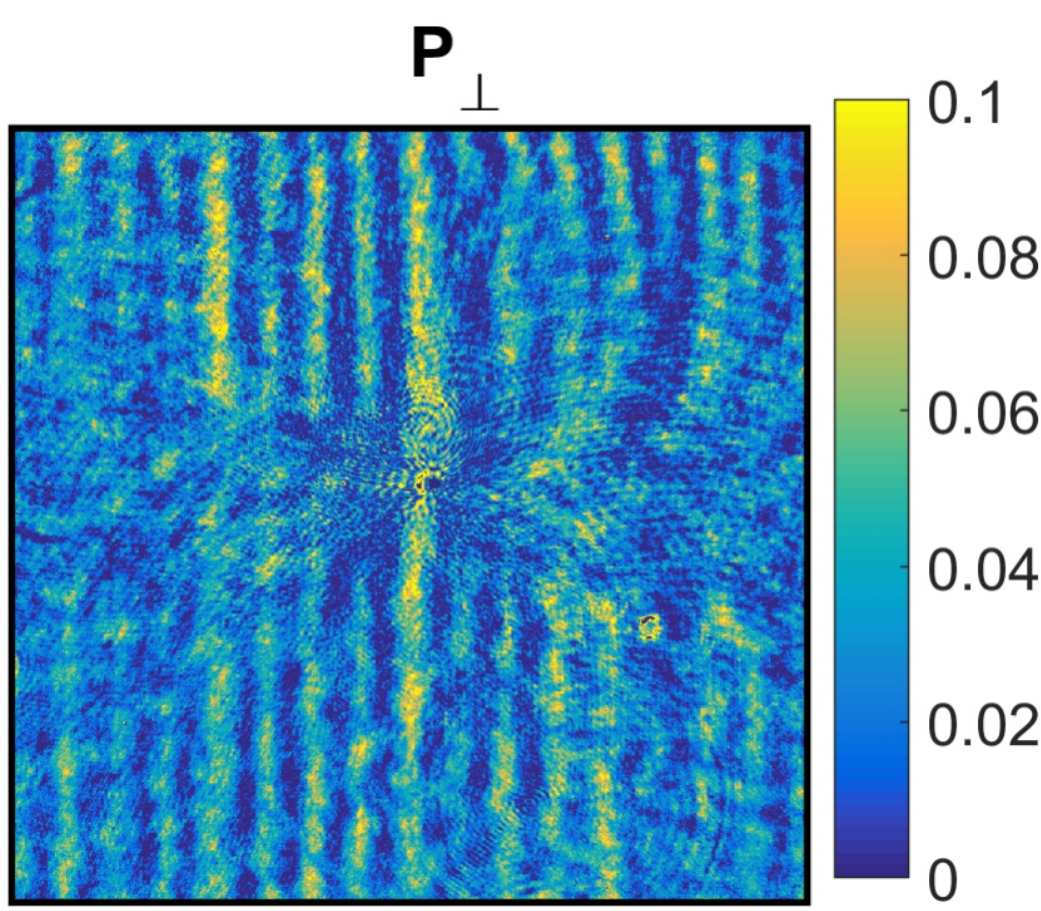} \\
				\includegraphics[width=0.49\columnwidth]{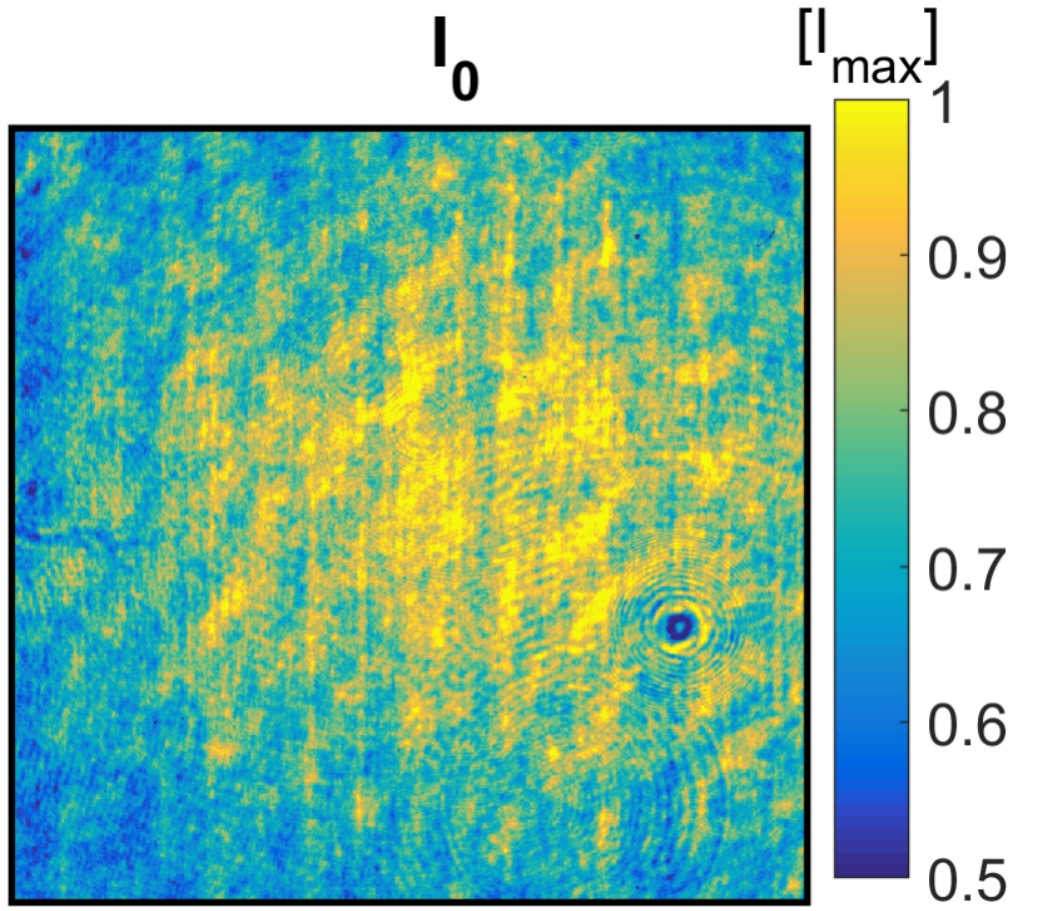}
				\includegraphics[width=0.49\columnwidth]{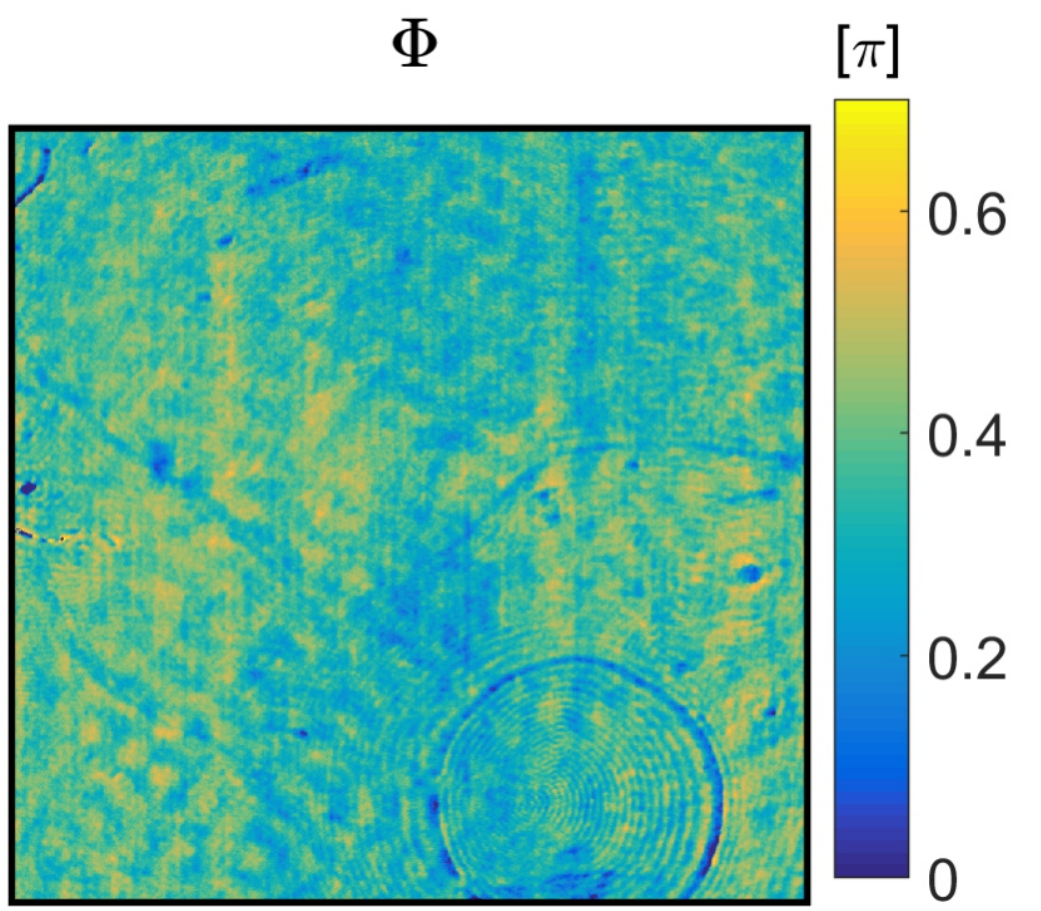}
			\caption{Measurement results of radial polarizer with a-PPSI (3/3C) algorithm: the orientation $\theta$, and transmission coefficients $P_{\parallel/\perp}$, mean intensity $I_0$ and phase front $\Phi$.}
			\label{fig:RES1a}
		\end{figure}

Figure \ref{fig:RES1a} presents the results of a-PPSI measurement of a radial polarizer made with a $(3/3B)$ algorithm from table \ref{tab:ALG2}.
				
This optical element generates an additional vortex phase term in form of a $2\pi$-spiral together with the desired polarization distribution by the outgoing light beam. There are various solutions cancelling this geometric phase term for the analysis of the generated polarization \cite{Rothau:14}. The a-PPSI procedure is analysing the pure object phase transmission $\Phi_{Obj}$ directly, so the phase result does not have this disturbing vortex term. 
				
The measured orientation $\theta$ of the transmission axis describes a full rotation similar to the azimuthal angle. The transmission coefficient $P_\parallel$ is nearly 1 and $P_\perp$ is very low, that fits to the expectations of the radial polarizer.

\section{Conclusion}
In this publication a further developed application of the novel method of the polarization and phase shifting interferometry (PPSI) is presented. With this interferometric method the simultaneous and full field measurement of arbitrary and spatially variant distributions of the phase transmission and the polarization dependent absorption of an object under test is possible.

Several different algorithms were introduced theoretically and two methods are used for the characterization of the specimen. The possible algorithms differ in the number of frames and variation steps and strategies. The minimal methods are working with just six or nine interferograms, depending on the type of dichroism. Furthermore, the full characterization behind the object regarding the polarization and phase of the outgoing light can be evaluated from the detected interferograms with e-PPSI calculation without any additional measurement.

The measurement setup is realized as a Mach-Zehnder interferometer for the observation of light fields in transmission. In principle, the setup may also be applied to measurements in reflected light,by changing it to a Michelson type setup.

\section{Funding Information} 	The authors thank the German Research Foundation (Deutsche Forschungsgemeinschaft, DFG) for funding this project under the support code (LI 1612/8-1).
\vspace*{3mm}
\begin{strip}
 \appendix 
    \section*{Appendix}	\setcounter{equation}{0}		\renewcommand{\theequation}{A{\arabic{equation}}}
Final intensity description in case of circular dichroism:
\begin{align}
\begin{split}
I=E_R^2+\frac{1}{2}E_{IN}^2&(P_r^2+P_l^2)+\frac{1}{2}E_{IN}^2(P_r^2-P_l^2)\sin\delta+\\
\frac{E_{IN}E_R}{\sqrt{2}}&\big[P_r\cos\Phi (\cos(\varphi-\alpha')-\sin(\varphi-\alpha'-\delta))+
	P_l\cos\Phi (\cos(\varphi+\alpha')+\sin (\varphi+\alpha'-\delta))+\\
	&P_r\sin\Phi(\sin(\varphi-\alpha')+\cos(\varphi-\alpha'-\delta))+P_l\sin\Phi (\sin(\varphi+\alpha')-\cos(\varphi+\alpha'-\delta))\big]
\label{eq:App2C}
\end{split}
\end{align}

Final intensity description in case of linear dichroism:
\begin{align}
\begin{split}
I=E_R^2+&E_{IN}^2[P_\parallel^2\cos^2(\theta-\alpha)+P_\perp^2\sin^2(\theta-\alpha)]+\\
&2E_{IN}E_R\cos(\Phi-\varphi)[P_\parallel\cos(\theta-\alpha')\cos(\theta-\alpha)+P_\perp\sin(\theta-\alpha')\sin(\theta-\alpha)]
\label{eq:App2L}
\end{split}
\end{align}
\end{strip}
\bibliographystyle{unsrt}  
\bibliography{sample}
\end{document}